# Infinite multiverses and where to find them?

Anubhav Kumar Srivastava[1], Pavel P. Popov[1], Guillem Müller-Rigat[2], and Maciej Lewenstein[1,3]

[1] ICFO-Institut de Ciencies Fotoniques, The Barcelona Institute of Science and Technology, Castelldefels (Barcelona) 08860, Spain.
[2] Faculty of Physics, Astronomy and Applied Computer Science, Jagiellonian University, ul. Łojasiewicza 11, 30-348 Kraków, Poland
[3] ICREA, Pg. Lluís Companys 23, 08010 Barcelona, Spain.

## ABSTRACT

Have you ever watched superhero movies like Spider-Man: Into the Spider-Verse? Or played games where your choices create different outcomes? What if we told you that in the real world, something even crazier might be happening all the time, right under our noses? Imagine shrinking down to the size of an atom. What you'd see wouldn't be like our everyday world at all! This is the realm of quantum physics, where the rules we know do not apply, where things exist everywhere and nowhere at once. The moment you observe something, it starts behaving differently. In this article, we will explore two of the many possible explanations for such phenomena, namely the Copenhagen interpretation and the many-worlds interpretation of quantum physics. We will also try to answer the question of whether there are many copies of you roaming around in different universes, and why you haven't met one.

## A JOURNEY INTO THE QUANTUM UNIVERSE: ARE THERE OTHER YOUs OUT THERE?

Have you ever wondered what the smallest thing is that makes up everything around us? Take the example of the screen or the page where you are reading this article. What is the smallest particle that makes this object? If your answer is atoms, you are somewhat correct, but what makes up even these indivisible particles? This question led to the discovery of even smaller sub-atomic particles – protons, electrons, and neutrons, which are further made of even smaller quarks, neutrinos, and gluons. These collectively form the standard model of particle physics along with some fundamental forces of nature. But do these tiny particles follow the same rules as we? No! Welcome to the quantum world! It's a place so strange, so bizarre, and so different from our everyday lives that the rules we're used to don't apply. This is the realm of quantum physics, the science that seeks to understand the remarkable behavior of the universe at its smallest possible scale.

    But this science comes with a huge, mind-boggling puzzle. It's a puzzle so deep that it has split physicists into different groups, each with its own incredible explanation for what reality is. However, these "interpretations" are not proven facts. They are helpful ways to make sense of what might be happening based on experiments and mathematics. The two most widely accepted explanations are the Copenhagen interpretation and the many-worlds interpretation. The latter suggests something straight out of science fiction: that there are endless parallel universes, each one containing a slightly different version of our world, our history, and even of you. We will dig deeper into this concept later in the article. Stay tuned!

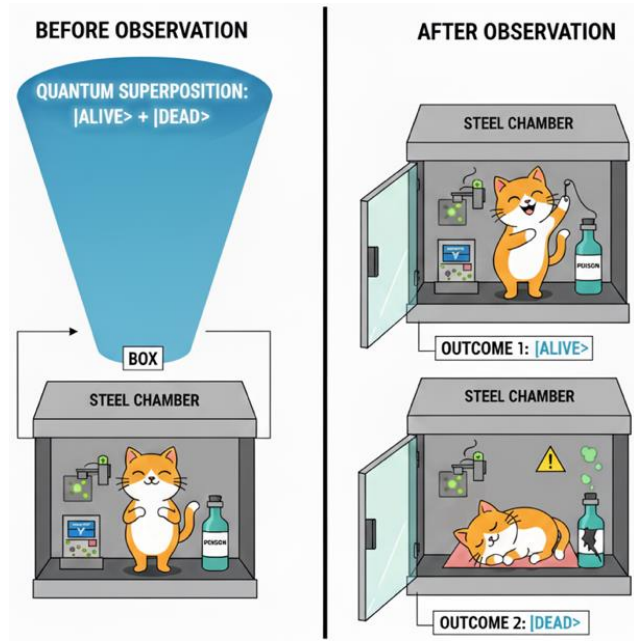

**Figure 1:** Schrödinger's cat experiment: The cat is in a box with a closed bottle of poison, which might or might not be released. We don't know if the cat is alive or dead until we measure by opening the box. Therefore, the cat is in a superposition of being both alive and dead at the same time before the box is opened. [Image generated by Google Gemini.]

## THE WEIRDNESS OF BEING SMALL: A WORLD OF MAYBES [1]

In our big, everyday world, things are defined and predictable. For example, when you kick a soccer ball, it either goes into the goal or it doesn't. There's no in-between! But in the tiny quantum world, things work very differently. A small particle, like an electron, can be in many places at once before we look at it. It is kind of "fuzzy" and not stuck in just one spot. Instead, it is in a mix of all its possible positions. Scientists call this strange state superposition, when something can be in several places at the same time. To help explain this weird idea, we must introduce one of the most eccentric pets in the world — Schrödinger's cat [2]. Erwin Schrödinger was a leading quantum physicist in the 1900s, and he proposed one of the most bizarre and perplexing thought experiments in quantum physics (also discussed in the FYM article: *The Awe-Inspiring Power of Quantum Computers*). In his story, there is a cat locked in a box with a special type of poison that might or might not be released. Because this is a random process, no one can know what happened to the cat until the box is opened. So, before opening the box, the cat is in a superposition state of being both dead and alive, just like the fuzzy electron is in many places at the same time. The moment we open the box, the cat's maybe-alive-or-maybe-dead state disappears, and it instantly picks one single reality, and the electron, which was previously in many places, is suddenly right here upon observing.

This sudden "collapse" into a single reality is the core puzzle of quantum mechanics. Why does it happen? And how does the particle or the cat choose which outcome to show us? The most common answer to this question has been, well, we don't know. In the following sections, we will look at two of the most popular interpretations that try to explain this mystery, and you can decide which one sounds more fascinating to you.

INTERPRETATION #1: THE COPENHAGEN "MIRACLE"

The first major attempt to explain this puzzle is called the Copenhagen interpretation [2–6]. As the name suggests, this interpretation of quantum physics was developed in Copenhagen, Denmark, by two of the 20th century's greatest scientists — Niels Bohr and Werner Heisenberg. It says that the act of measurement itself forces the universe to make a choice. The fuzzy cloud of possibilities of being in different states at the same time, which scientists call the wavefunction, is said to "collapse" into a single definitive state, as seen in the case of the electron and Schrödinger's cat. This is the most widely accepted interpretation of quantum physics in the scientific community due to its simplicity and practicality. It beautifully connects the classical and quantum worlds and offers a solid mathematical framework for understanding these quantum systems. However, it doesn't say how or why this collapse happens. It just does. For some physicists, this is not a digestible argument that something happens, and we have to accept it without further questions. Science has no room for miracles, and this "wavefunction collapse" felt a little too magical for comfort. This dissatisfaction led some scientists to search for a better, more complete explanation, even if it led to a truly astonishing conclusion.

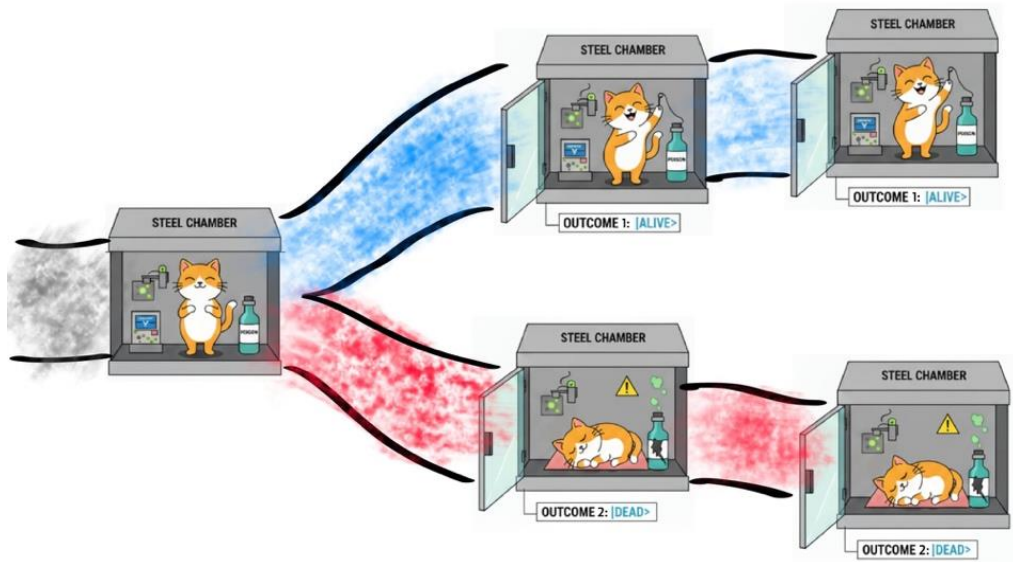

**Figure 2:** Many-worlds interpretation: The universe splits into two realities every time an action or a choice is made. Here, one branch of reality is where the poison was not released, and the cat is alive and happy. The other branch is where the poison was released, and the cat died. When we measure the state of the cat in the box, we restrict ourselves to one of the two branches of reality.

INTERPRETATION #2: THE MANY-WORLDS SOLUTION

In 1957, a young physicist named Hugh Everett III proposed a radical and brilliant solution to the collapse problem [2, 6, 7]. He asked a simple question: What if the wavefunction never collapses at all? What if, instead of the universe picking one reality, all possible realities happen simultaneously? Say you have to choose between pizza and pasta tonight. According to the many-worlds view, if something is possible, then it must occur in some branch of the universe. Therefore, there is a universe where you chose a pizza for dinner, and in that universe, you're happily munching on pizza toppings. However, in a parallel universe, a different you is twirling spaghetti! Every quantum event creates a new branch on a giant cosmic tree. Each branch is a new universe, where a different possibility became real, as shown in Figure 2, where the state of the cat splits into two branches – one where the cat ate the poison, and the other where it is still alive and happy.

This isn't just a fun idea; it elegantly addresses the "miracle" problem. There is no mysterious collapse. The wavefunction, the quantum rulebook, evolves smoothly, creating more and more branches as time goes on. In fact, by choosing pizza over spaghetti, you have created your own branch of the universe! Isn't that cool and weird at the same time? So, have we solved the quantum mystery? Can we dust off our hands and go home? No, we have given one possible solution, but raised ten other questions. This complicated interpretation also faces widespread criticism among the physicists, as trillions of branches are created every instant. It also raises the philosophical question of which "you" is actually you and how come you are here in this exact universe and not elsewhere. If these other universes are real, where are they, and why have you not met another version of you? The answer lies in a critical physical process called decoherence.

## THE GREAT DIVIDE: WHAT IS DECOHERENCE?

Decoherence is the process that seals these parallel universes off from one another, making them completely inaccessible to each other. Imagine two identical dogs starting a race along two identical paths in opposite directions around a huge park [1]. The first dog runs without stopping towards the finish line. However, the second dog faces various distractions on its path – squirrels, dropped food, and even rabbits. These interactions affect the speed and path of the second dog, and it is no longer in sync with the first. They no longer meet at the finish line at the same time.

Decoherence is like the universe's collection of squirrels and dropped food. The different branches of the universe start out connected. If you could bring them back together, they would interfere with each other, proving they are part of the same system. But our universe is a messy place, ensuring that each independent branch of reality runs its own separate race, constantly interacting in tiny, random ways with its environment – bumping into air molecules, photons, and everything else. That's why we only ever experience our own single branch of the universe. Hence, decoherence provides a possible solution for the collapse problem [2] and the classical nature of reality around us.

## SO, WHICH IDEA IS RIGHT AND WHY DO WE CARE?

The incredible thing is that we have no way of knowing which interpretation is correct. No known experiment can distinguish between a universe that collapses (Copenhagen) and one that splits (many-worlds). Both explanations predict the same outcomes for all the experiments we can actually perform.

It is your personal "taste" whether you prefer a single universe where reality makes a mysterious, magical leap every time we look at it, or you prefer a universe that follows smooth, unbroken mathematical rules, but at the cost of creating an infinite number of parallel worlds with every quantum flicker?

Remember, the next time you're faced with a choice, somewhere out there, in a parallel universe, another you might have picked the other option! The universe is a lot stranger and more wonderful than we often imagine. Keep exploring, keep asking questions, and maybe one day you'll be the one to unlock the next big secret of the multiverse!

ACKNOWLEDGEMENTS

We thank Harjasnoor Kakkar for insightful discussions. AKS acknowledges support from the European Union's Horizon 2020 Research and Innovation Programme under the Marie Sklodowska-Curie Grant Agreement No. 847517. GMR acknowledges financial support from the European Union under ERC Advanced Grant TAtypic, Project No. 101142236. The ICFO group acknowledges support from MCIN/AEI (PGC2018–0910.13039/501100011033, CEX2019-000910-S/10.13039/501100011033, Plan National STAMEENA PID2022-139099NB, project funded by MCIN and by the "European Union NextGenerationEU/PRTR" (PRTR-C17.I1), FPI); Ministry for Digital Transformation and of Civil Service of the Spanish Government through the QUANTUM ENIA project call - Quantum Spain project, and by the European Union through the Recovery, Transformation and Resilience Plan – NextGenerationEU within the framework of the Digital Spain 2026 Agenda; CEX2024-001490-S [MICIU/AEI/10.13039/501100011033]; Fundació Cellex; Fundació Mir-Puig; Generalitat de Catalunya (European Social Fund FEDER and CERCA program); Barcelona Supercomputing Center MareNostrum (FI-2023-3-0024); European Union HORIZON-CL4-2022-QUANTUM-02-SGA – PASQuanS2.1, 101113690; EU Horizon 2020 FET-OPEN OPTOlogic, Grant No 899794; QU-ATTO, 101168628; EU Horizon Europe research and innovation program under grant agreement No. 101080086 NeQST.

AI TOOL STATEMENT

The authors declare that Gen AI was used in the creation of this manuscript. Figure 1 was generated using Google Gemini, and Figure 2 was modified by the authors from Figure 1.